\documentclass[conference]{IEEEtran}
\usepackage{amsmath,amssymb,amsfonts}
\usepackage{graphicx}
\usepackage{textcomp}
\usepackage{xcolor}

\usepackage{xcolor}
\usepackage[linesnumbered,ruled,vlined]{algorithm2e}
\usepackage{listings}
\usepackage[noend]{algpseudocode}

\SetCommentSty{mycommfont}
\SetKwInput{KwInput}{Input}   % Set the Input
\SetKwInput{KwOutput}{Output} % set the Output
\usepackage{subcaption}
\usepackage{caption}

\definecolor{GreenForest}{rgb}{0.09, 0.45, 0.27}
\usepackage{soul}
\usepackage{balance}
\usepackage{cite}
\usepackage{acro} 
\DeclareAcronym{RIS}{
  short = RIS ,
  long  = reconfigurable intelligent surfaces ,
  class = abbrev
}
\DeclareAcronym{6G}{
  short = 6G,
  long  = 6th generation ,
  class = abbrev
}
\DeclareAcronym{5G}{
  short = 5G,
  long  = 5th generation ,
  class = abbrev
}
\DeclareAcronym{EM}{
  short = EM ,
  long  = electromagnetic,
  class = abbrev
}
\DeclareAcronym{BS}{
  short = BS,
  long  = base station ,
  class = abbrev
}
\DeclareAcronym{UE}{
  short = UE ,
  long  = user equipment ,
  class = abbrev
}
\DeclareAcronym{MISO}{
  short = MISO,
  long  = multiple-input single-output ,
  class = abbrev
}
\DeclareAcronym{MMSE}{
  short = MMSE ,
  long  = minimum mean squared error ,
  class = abbrev
}
\DeclareAcronym{DFT}{
  short = DFT,
  long  = discrete Fourier transform ,
  class = abbrev
}
\DeclareAcronym{THz}{
  short = THz,
  long  = Terahertz ,
  class = abbrev
}
\DeclareAcronym{IoT}{
  short = IoT,
  long  = internet of things ,
  class = abbrev
}
\DeclareAcronym{MSE}{
  short = MSE,
  long  = mean square error ,
  class = abbrev
}
\DeclareAcronym{CSI}{
  short = CSI ,
  long  = channel state information ,
  class = abbrev
}
\DeclareAcronym{MIMO}{
  short = MIMO,
  long  = multiple-input multiple-output ,
  class = abbrev
}
\DeclareAcronym{UPA}{
  short = UPA,
  long  = uniform planner array ,
  class = abbrev
}
\DeclareAcronym{RF}{
  short = RF,
  long  = radio-frequency ,
  class = abbrev
}
\DeclareAcronym{mmWave}{
  short = mmWave,
  long  = millimeter-wave ,
  class = abbrev
}
\DeclareAcronym{AoA}{
  short = AoA ,
  long  = angle of arrival ,
  class = abbrev
}
\DeclareAcronym{AoD}{
  short = AoD,
  long  = angle of departure ,
  class = abbrev
}
\DeclareAcronym{EKF}{
  short = EKF,
  long  = extended Kalman filter ,
  class = abbrev
}
\DeclareAcronym{LMS}{
  short = LMS,
  long  = least mean square ,
  class = abbrev
}
\DeclareAcronym{BiLMS}{
  short = BiLMS,
  long  = bi-directional LMS ,
  class = abbrev
}
\DeclareAcronym{SNR}{
  short = SNR,
  long  = signal-to-noise ratio ,
  class = abbrev
}
\DeclareAcronym{LoS}{
  short = LoS,
  long  = line-of-sight ,
  class = abbrev
}
\DeclareAcronym{TDD}{
  short = TDD,
  long  = time-division duplexing ,
  class = abbrev
}
\DeclareAcronym{NMSE}{
  short = NMSE,
  long  = normalized mean square error ,
  class = abbrev
}
\DeclareAcronym{SDR}{
  short = SDR,
  long  = semidefinite relaxation ,
  class = abbrev
}
\DeclareAcronym{QoS}{
  short = QoS,
  long  = quality of service ,
  class = abbrev
}
\DeclareAcronym{NOMA}{
  short = NOMA,
  long  = non-orthogonal multiple access ,
  class = abbrev
}
\DeclareAcronym{OMA}{
  short = OMA,
  long  = orthogonal multiple access ,
  class = abbrev
}
\DeclareAcronym{NU}{
  short = NU,
  long  = near user ,
  class = abbrev
}
\DeclareAcronym{FU}{
  short = FU,
  long  = far user ,
  class = abbrev
}
\DeclareAcronym{SIC}{
  short = SIC,
  long  = successive interference cancellation ,
  class = abbrev
}
\DeclareAcronym{PLS}{
  short = PLS,
  long  = physical layer security ,
  class = abbrev
}
\DeclareAcronym{MRT}{
  short = MRT,
  long  = maximum ratio transmission ,
  class = abbrev
}
\DeclareAcronym{AWGN}{
  short = AWGN,
  long  = additive white Gaussian noise,
  class = abbrev
}
\DeclareAcronym{SINR}{
  short = SINR,
  long  = signal-to-interference-plus-noise ratio ,
  class = abbrev
}
\DeclareAcronym{BPSK}{
  short = BPSK,
  long  = binary phase shift keying ,
  class = abbrev
}
\DeclareAcronym{QPSK}{
  short = QPSK,
  long  = quadrature phase shift keying ,
  class = abbrev
}
\DeclareAcronym{SVD}{
  short = SVD,
  long  = singular value decomposition ,
  class = abbrev
}
\DeclareAcronym{EVD}{
  short = EVD,
  long  = eigenvalue decomposition ,
  class = abbrev
}
\DeclareAcronym{PDF}{
  short = PDF,
  long  = probability density function ,
  class = abbrev
}
\DeclareAcronym{SER}{
  short = SER,
  long  = symbol error rate ,
  class = abbrev
}
\DeclareAcronym{MGF}{
  short = MGF,
  long  = moment generating function ,
  class = abbrev
}
\DeclareAcronym{2D}{
  short = 2D,
  long  = two-dimensional ,
  class = abbrev
}
\DeclareAcronym{3D}{
  short = 3D,
  long  = three-dimensional ,
  class = abbrev
}
\DeclareAcronym{CLT}{
  short = CLT,
  long  = central limit theorem ,
  class = abbrev
}
\DeclareAcronym{QAM}{
  short = QAM,
  long  = quadrature amplitude modulation ,
  class = abbrev
}
\DeclareAcronym{SISO}{
  short = SISO,
  long  = single-input single-output ,
  class = abbrev
}
\DeclareAcronym{CE}{
  short = CE,
  long  = channel estimation ,
  class = abbrev
}
\DeclareAcronym{KG}{
  short = $K_G$,
  long  = generalized-K ,
  class = abbrev
}
\DeclareAcronym{LSKRF}{
  short = LSKRF,
  long  = least squares Khatri-Rao factorization ,
  class = abbrev
}

\begin{document}
\bstctlcite{IEEEexample:BSTcontrol}
\title{Reconfigurable intelligent surface (RIS): Eigenvalue Decomposition-Based Separate Channel Estimation}

\author { \IEEEauthorblockN{Salah Eddine Zegrar\IEEEauthorrefmark{1}, Liza Afeef\IEEEauthorrefmark{1}, and H\"{u}seyin Arslan\IEEEauthorrefmark{2}}
   
    \IEEEauthorblockA{\IEEEauthorrefmark{1}Department of Electrical and Electronics Engineering, Istanbul Medipol University,\\
     Istanbul, Turkey (e-mail:
    \{salah.zegrar@std.medipol.edu.tr, liza.shehab@std.medipol.edu.tr,  huseyinarslan@medipol.edu.tr\})}
   
    \IEEEauthorblockA{\IEEEauthorrefmark{2}Department of Electrical Engineering, University
of South Florida,\\ Tampa, FL, USA,(e-mail:~arslan@usf.edu)
        }}
% ====================================================================
\maketitle

% === ABSTRACT =====================================
% ==========================================================================
\begin{abstract}
Reconfigurable intelligent surface (RIS) has recently drawn significant attention in wireless communication technologies. However, identifying, modeling, and estimating the RIS channel in multiple-input multiple-output (MIMO) systems are considered challenging in recent studies. In this paper, a disassembled channel estimation framework for the RIS-MIMO system is proposed based on the eigenvalue decomposition (EVD) concept to separate the cascaded channel links and estimate each link separately. This estimation is based on modeling the RIS-MIMO channel as a keyhole MIMO system model. Numerical results show that the proposed estimation method has a low estimation time overhead while providing less estimation error.
\footnote{This work has been submitted to the IEEE for possible publication. Copyright may be transferred without notice, after which this version may no longer be accessible.}

\end{abstract}

% === KEYWORDS ====================================================================
% =================================================================================
\begin{IEEEkeywords}
Keyhole channel, reconfigurable intelligent surface, multiple-input multiple-output, eigenvalue decomposition.
\end{IEEEkeywords}

\IEEEpeerreviewmaketitle

% === I. INTRODUCTION =============================================================
% =================================================================================
\section{Introduction}
Recently, \ac{RIS}s have been studied as a highly potential technology that can face the service requirements of the \ac{6G} wireless networks and beyond. \ac{RIS}’s capability arises from the ability to control and change the wireless channel from a highly time-varying to a deterministic one. The \ac{RIS} elements can steer the reflected electromagnetic wave toward any specific direction with accurate angle \cite{Survey-ertugrul}, this provides a great potential in enhancing the system's performance and security \cite{furqan-survey}. Hence, the \ac{RIS}-based wireless transmission can have a great potential for realizing \ac{MIMO} technologies. However, many challenges in these \ac{RIS} systems are raised up to the field such as channel estimation problem.

Channel estimation in \ac{RIS}-aided networks is very critical, since the real time applicability of the \ac{RIS} is proportionally related to the resolution and the time overhead of the channel estimation. For instance, the authors in \cite{wang-channel} proposes a three-phase channel estimation algorithm for the uplink \ac{RIS}-assisted \ac{MISO} system. In the first phase, the direct channels between \ac{UE}s-\ac{BS} are estimated while the \ac{RIS} elements are turned off. In the following phase, only one \ac{UE} transmits the pilot signal and the cascaded channel is estimated only for it. In the last phase, only the scaling factors need to be estimated since the channel between all \ac{UE}s and the \ac{RIS} is considered correlated. However, the training overhead, the number of supportable users, and the performance of channel estimation are the limits of this technique.
In the approach of \cite{OFDM}, the \ac{RIS} elements are divided into sub-surfaces while assuming full reflected power during the channel estimation and data transmission. Each sub-surface consists of $M$ adjacent elements with a common reflection coefficient to minimize the complexity of the system.
In \cite{unbiased}, a \ac{DFT}-based channel estimation method is proposed, where all the \ac{RIS} elements are switched on during the whole channel estimation period, and the \ac{DFT} matrix is used to determine the phases of these elements.
Parallel factor-based channel estimation was proposed in \cite{wang-channel,parafac1,parafac2}, where the cascaded channel is unfolded by decomposing the \ac{3D} representation of the received signal using eigen decomposition. Although they obtain the channels \ac{BS}-\ac{RIS} and \ac{RIS}-\ac{UE} separately, the algorithm has high complexity and time overhead.
While the authors in \cite{zegrar} estimated the channels \ac{BS}-\ac{RIS} and \ac{RIS}-\ac{UE} separately and emphasized the importance of the separate estimation, they also presented a new algorithm to track mobile users communicating throughout \ac{RIS}.\\

The aforementioned channel estimation works in \cite{wang-channel,OFDM,unbiased,parafac1} consider the overall concatenated effective channel estimation, where this type of estimation leads to channel statistics loss compared to estimating RIS-MIMO channels separately. The importance raises from the fact that the RIS has the ability to precode the incident signal if both Tx-\ac{RIS} and \ac{RIS}-Rx \ac{CSI} are available. For instance, if the cascaded \ac{CSI} is known to the transmitter, the reflection coefficient of the RIS elements can be optimized to perform all kind of multiple-antenna precoding schemes, such as zero forcing the Tx-\ac{RIS} channel to control the \ac{RIS}-Rx independently. Besides, estimating both channels separately allows us to identify the behavior of the channel in each part whether it is a time-varying or time-invariant channel, and thus enable channel tracking by setting the phases accordingly, as discussed in detail in \cite{zegrar}. Therefore, a generic representation of the channel is needed, that can allow us to analyse the \ac{RIS}-aided systems in more details and under various assumptions and conditions to have more insight about the composition of the channel.

For the cascaded channel representation for the RIS-MIMO systems, the work in \cite{parafac2} presents a channel estimation method based on parallel factor decomposition algorithm to unfold the cascaded channel Tx-\ac{RIS} and \ac{RIS}-Rx by decomposing the \ac{3D} representation of the received signal using eigen decomposition. However, these methods suffer from high complexity due to the three dimensional matrix operations which eventually increases the channel estimation time overhead. Another solution to reduce the pilot overhead is to use the sparse matrix factorization and matrix completion method if the channel exhibits the low-rank property \cite{cascaded}.

Motivated by the above discussion, in this paper, a low-complex channel estimation algorithm is proposed based on the keyhole channel and \ac{EVD} concept that enables the estimation for each channel in the RIS-MIMO system separately. Estimating the cascaded channels separately in the RIS-aided systems is a key enabler to exploit full \ac{RIS} power. For instance, serving mobile \ac{UE}s via \ac{RIS} is still a large gap in the literature since it is almost impossible to determine whether the time-varying channel is caused by any changes in the environment between Tx and \ac{RIS} or \ac{RIS} and Rx or due to the mobility of the \ac{UE}.

The contributions of this work are summarized as follows:
\begin{itemize}
    \item Disassembling low-complex channel estimation algorithms are proposed based on the \ac{EVD} concept and modeling the \ac{RIS}-\ac{MIMO}-based channel as a keyhole \ac{MIMO} channel. The proposed algorithm unfolds the cascaded \ac{RIS}-\ac{MIMO} channel and estimates each channel part separately due to the fact that each \ac{RIS} element generates rank-one channel matrix.
    \item The numerical analysis validates the proposed algorithms' performance gains in terms of reduced estimation time overhead compared to conventional methods. 
\end{itemize}

% =================================================================================
\section{RIS Channel Modelling}\label{section:Channel-model}

\begin{figure}[t]
    \begin{center}
    \subfloat[]{\includegraphics[scale=0.33]{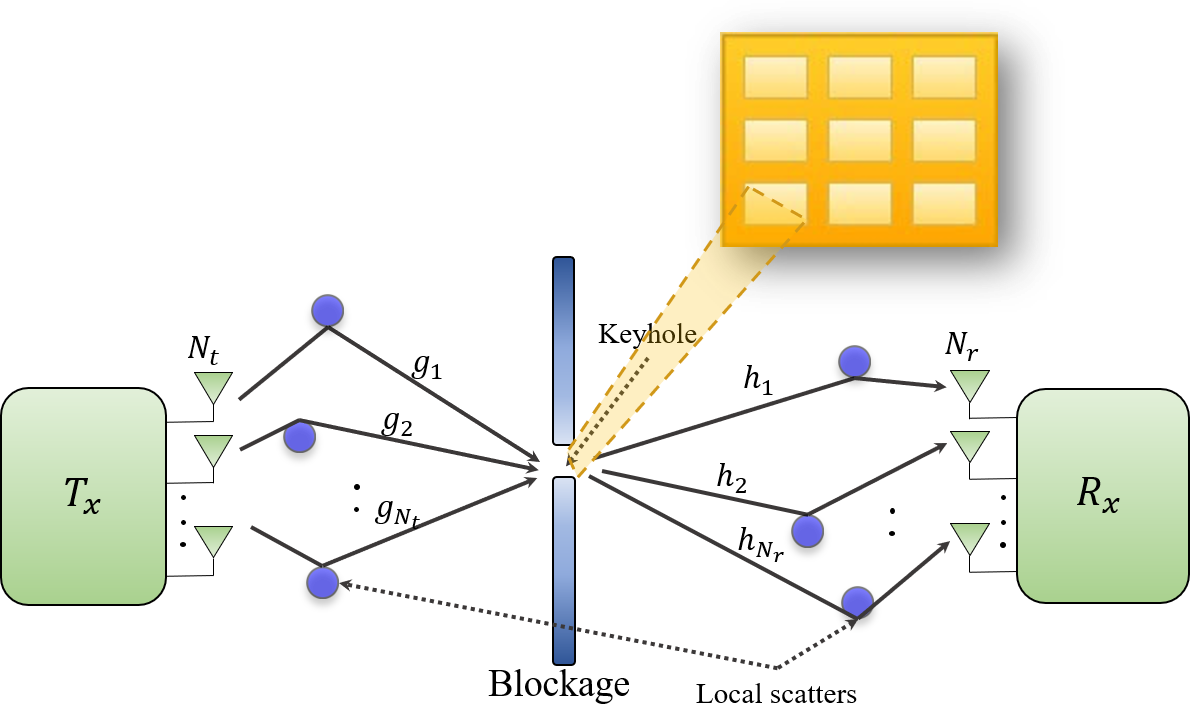}}
    \hspace{0.9cm}
    \subfloat[]{\includegraphics[scale=0.33]{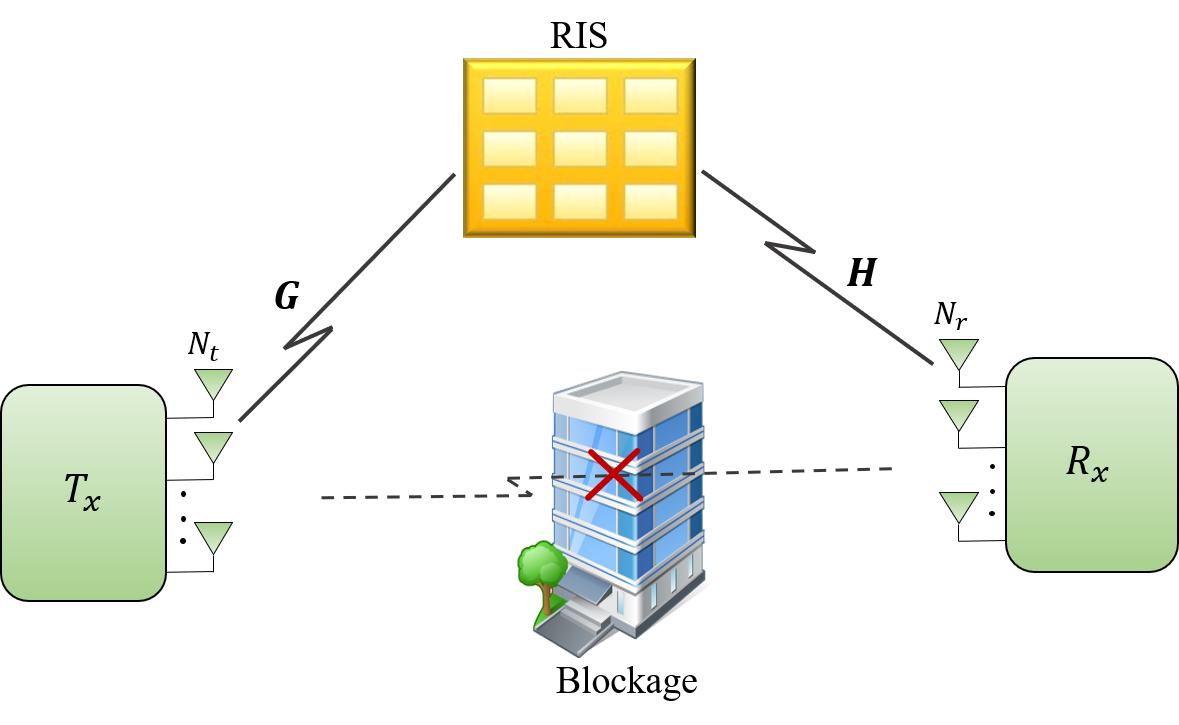}}
    \end{center}
    \centering
    \caption{(a) A single-keyhole \ac{MIMO} system equivalent to single-element \ac{RIS}-\ac{MIMO} system, and (b) \ac{RIS}-\ac{MIMO} system model.}
    \label{fig:system-model_all}
\end{figure}

In this section, the channel model for \ac{RIS} is derived based on the concept of keyholes. \ac{RIS} elements are assumed to be regular scatterers with an ability to control the phases of the scattered signal \cite{Survey-ertugrul}, and the channel model is developed accordingly in this paper. Also, the \ac{RIS} is assumed to be operating in the far field.

%++++++++++++++++++++++++++++++++++++++++++++++++++++++++++
\subsection{Keyhole \ac{MIMO} Channel Model} \label{subsection:keyhole}

In practice, a \ac{MIMO} system can also operate in an insufficient scattering environment, which can lead to a rank deficient channel. For example, in a scenario of transmitter and receiver surrounding by clutters (rich scattering environment), where most of the signal energy passes through small holes, the channel rank can reduce to one. This effect has been called "keyhole", and the channel is modeled as a keyhole \ac{MIMO} channel \cite{keyhole2}. It can happen when the band of scatterers around the transmitter and the receiver is small compared to the distance between the transmitter and the receiver. Other scenarios for the keyhole effects can be given in \cite{ngo2017no}.    
Following similar behaviour, it is proved in \cite{keyhole_RIS_stolenwork} that the \ac{RIS}-\ac{MIMO} channel model is similar to keyhole \ac{MIMO} channels by deriving a closed-form approximation to the channel distribution of the \ac{RIS}-aided systems which is appeared to be equivalent to keyhole channels under some limitations.

The concept of keyhole \ac{MIMO} channel was firstly introduced in \cite{Keyhole_original}, where the channel is considered as a dyad with one degree of freedom which usually appears in a general \ac{MIMO} relay channel \cite{keyhole_relay}. In a single-keyhole \ac{MIMO} system model as shown in Fig. \ref{fig:system-model_all}(a), the total channel matrix can be given as
\begin{equation}
\begin{aligned}
\mathbf{H}_{\mathrm{key}}&=
  \begin{pmatrix}
h_1\\
\vdots\\
h_{N_r} 
 \end{pmatrix}
 \sigma_{scs}
   \begin{pmatrix}
g_1 & \cdots &g_{N_t} \\
 \end{pmatrix},\\
\mathbf{H}_{\mathrm{key}}& =\sigma_{scs}
 \begin{pmatrix}
 h_1 g_1  & \cdots& h_1 g_{N_t}  \\
\vdots& \ddots & \vdots\\
  h_{N_r} g_1 & \cdots& h_{N_r}g_{N_t}
 \end{pmatrix},
 \end{aligned}
 \label{equ:Keyhole-matrix}
\end{equation}    
where $\sigma_{scs}\in[0,1]$ is the scattering cross-section of the keyhole, $g_1$ and $g_2$ are the complex channel coefficients of the Tx-keyhole channel, $h_1$ and $h_2$ are the channel coefficients of the keyhole-Rx channel.
Clearly, the entries of $\mathbf{H}_{\mathrm{key}}$ are uncorrelated; however, this matrix has only one degree of freedom unlike the case without keyhole (\ac{MIMO} channel) and hence it has low capacity. Therefore, the \ac{SVD} of $\mathbf{H_{\mathrm{key}}}$ is written as
\begin{equation}
    \mathbf{H}_{\mathrm{key}} = \mathbf{U}\Lambda\mathbf{V}^H=\sum_{m=1}^{\min(N_t,N_r)} \mathbf{u}_m\lambda_m\mathbf{v}_m^H=\mathbf{u}_1\lambda_1\mathbf{v}_1^H,
    \label{equ:EVD}
\end{equation}
where $N_t$ and $N_r$ are the number of transmit and receive antennas, respectively. This means that the keyhole channel can be represented by only one dyad.

\subsection{RIS as a Keyhole} \label{subsection:RIS-keyhole}
From the discussion in Subsection \ref{subsection:keyhole}, we can observe that each \ac{RIS} element is a perfect example of keyhole. This has been expressed in \cite{keyhole_RIS_stolenwork}, where the authors showed the similarity between their derived \ac{RIS}-\ac{MIMO} channel distribution and the keyhole channels distribution.    

Considering a \ac{MIMO} system with $N_t$ transmit and $N_r$ receive antennas at the Tx and the Rx, respectively. The direct link Tx-Rx is ignored due to the bad propagation conditions. Therefore, an \ac{RIS} consisting of $N$ elements is deployed to assist the communication. It is assumed that the transmission operates in \ac{TDD} mode, and local scatterers are randomly distributed quasistatically near the transmit or the received antenna array. Thus, the channel becomes quasistatic, frequency flat, and uncorrelated.

Let $\mathbf{H}^i_T\in \mathbb{C}^{N_{r}\times N_{t}}$ be the total effective channel between Tx-Rx via the $i$-th \ac{RIS} element, then given \eqref{equ:Keyhole-matrix}, the total channel can be defined as
\begin{equation}
\begin{aligned}
 &\mathbf{H}_{T}^i=  \mathbf{h}^i\sigma_{scs}^i\gamma^i e^{j\theta_i}\mathbf{g}^i,
 \label{equ:single-element-matrix}
 \end{aligned}
\end{equation}
where $\theta_i$ is the controlled \ac{RIS} element phase, $\gamma^i$ and $\sigma_{scs}^i$ denote the gain and the scattering cross-section of the $i$-th element of the \ac{RIS}, respectively. $\mathbf{g}^i \in \mathbb{C}^{1\times N_{t} }$ and $\mathbf{h}^i \in \mathbb{C}^{N_{r}\times 1 }$ denote the Tx-$i$-th RIS element and the $i$-th RIS element-Rx channel vectors, respectively. It should be noted that all elements in matrix $\mathbf{H}_{T}^i$ are uncorrelated and $\mathbf{h}^i$ and $\mathbf{g}^i$ are independent, yet, $\mathbf{H}_{T}^i$ has a single degree of freedom $\operatorname{rank}(\mathbf{H}_{T}^i)=1$.
\footnote{\textbf{Remark 1.} \textit{It should be noted that if $N>\operatorname{max}(N_t,N_r)$, then the total effective channel $\mathbf{H}_{T}$ is a full rank matrix i.e., $\mathrm{rank}(\mathbf{H}_{T})=\min(N_t,N_r)$. This reflects the high capacity achieved by the \ac{MIMO} system when deploying \ac{RIS}}.}

The total Tx-\ac{RIS}-Rx channel is the sum of the contributions from each \ac{RIS} element. Therefore, for an \ac{RIS}-\ac{MIMO} channel model, the total effective channel is expressed as
\begin{equation}
    \mathbf{H}_{T} = \sum_{i=1}^N \mathbf{H}_{T}^i = \sum_{i=1}^N \mathbf{h}^i\sigma_{scs}^i\gamma^i e^{j\theta_i}\mathbf{g}^i = \mathbf{{H}} \mathbf{\Theta} \mathbf{{G}},
\end{equation}
where $\mathbf{H} = [\mathbf{h}^1,\dots,\mathbf{h}^N]$, $\mathbf{{G}} = [(\mathbf{g}^1)^T,\dots,(\mathbf{g}^N)^T]^T$ and $\mathbf{\Theta} = \operatorname{diag}([\sigma_{scs}^1 \gamma^1e^{j\theta_1},\dots,\sigma_{scs}^N\gamma^N e^{j\theta_N}])$ is the diagonal matrix containing the reflection coefficient induced by each element along with its gain and radar cross-section. Throughout this paper, it is considered that $\sigma_{scs}^i=\gamma^i=1$, for $\forall i$.

% ===========================================================
\section{Channel Estimation} \label{section:estimation}
To develop the channel estimation algorithm, firstly, the model is derived for only a single \ac{RIS} element. Then, it is generalized for the whole \ac{RIS}.
%++++++++++++++
\subsection{Channel Estimation for Single \ac{RIS} Element} \label{subsection:single-RIS}
We consider the $i$-th element of the \ac{RIS} to be activated, and all other elements to be in the off-mode. The same system model shown in Fig. \ref{fig:system-model_all}(b) is adopted. The Tx-\ac{RIS}-Rx channel is given by \eqref{equ:single-element-matrix}. Using the results concluded in Subsection \ref{subsection:keyhole}, the \ac{SVD} of the single \ac{RIS} element channel matrix is expressed as
\begin{equation}
    \mathbf{H}_{T}^i = \mathbf{U}^i\Lambda^i(\mathbf{V}^{i})^H=\mathbf{u}_1^i\lambda_1(\mathbf{v}_1^{i})^H.
    \label{equ:Channel-est-sing}
\end{equation}

By comparing (5) to (3), it can be seen that $\mathbf{u}_1^i$ and $(\mathbf{v}_1^{i})^H$ are normalized and rotated versions of the vectors $\mathbf{h}^i$ and $\mathbf{g}^i$, respectively, i.e., $\mathbf{h}^i = \sqrt{\lambda_1}\mathbf{u}_1^i e^{j\alpha_i}$ and $\mathbf{g}^i = \sqrt{\lambda_1}(\mathbf{v}_1^{i})^H e^{j\beta_i}$, where $e^{\alpha_i}$ and $e^{j\beta_i}$ ensures that $\mathbf{h}^i$ and $\mathbf{g}^i$ are not necessarily orthogonal. It is should be noted that $e^{j\alpha_i}=e^{-j\beta_i}$ therefore, their effect cancels out when multiplied, and estimating channel $\mathbf{H}_{T}^i$ is equivalent to estimating $\mathbf{u}_1^i\lambda_1(\mathbf{v}_1^{i})^H$.

Let $\mathbf{X} = [\mathbf{x}_1,\dots,\mathbf{x}_{N_t}]\in \mathbb{C}^{N_t\times N_t}$ be the transmitted pilot matrix over the total estimation time slots. It is preferable to design $\mathbf{X}$ and $\mathbf{\Theta}$ to be semi-unitary matrices i.e., $\mathbf{X}\mathbf{X}^H=\mathbf{I}$. Then, the received noisy signal ${\mathbf{Y}} \in \mathbb{C}^{N_r\times N_t}$ is given by
\begin{equation}
    {\mathbf{Y}} = \widetilde{\mathbf{Y}} + \mathbf{N}, 
    \label{equ:Rx}
\end{equation}
where $\widetilde{\mathbf{Y}} = \mathbf{H}_{T}^i \mathbf{X}$ is the received noise-free signal and $\mathbf{N}\in\mathbb{C}^{N_r\times N_t} $ is the zero-mean \ac{AWGN} with variance $\sigma^2$ .i.e, $\mathbf{N} \sim  \mathbb{C}\mathcal{N}(\mathbf{0}, \sigma^2 \mathbf{I})$. 

In order to estimate the cascaded channels separately, the proposed algorithm considers two cases as follows 
\begin{equation}
\begin{aligned}
   \mathbf{Y}\mathbf{Y}^H &= \mathbf{H}_{T}^i \mathbf{X}\mathbf{X}^H(\mathbf{H}_{T}^{i})^H + \widehat{\mathbf{N}}_1, \\
    & = \mathbf{U}^i\Lambda^i(\mathbf{V}^{i})^H\mathbf{X}\mathbf{X}^H\mathbf{V}^{i}\Lambda^i(\mathbf{U}^{i})^H + \widehat{\mathbf{N}}_1,
    \label{equ:RX2}
\end{aligned}
\end{equation} 

\begin{equation}
\begin{aligned}
     \mathbf{X} \mathbf{Y}^H&\mathbf{Y}\mathbf{X}^H = \mathbf{X}\mathbf{X}^H(\mathbf{H}_{T}^{i})^H\mathbf{H}_{T}^i \mathbf{X}\mathbf{X}^H + \widetilde{\mathbf{N}}_1, \\
    & = \mathbf{X}\mathbf{X}^H\mathbf{V}^{i}\Lambda^i(\mathbf{U}^{i})^H\mathbf{U}^i\Lambda^i(\mathbf{V}^{i})^H\mathbf{X}\mathbf{X}^H + \widehat{\mathbf{N}}_2,
    \label{equ:RX2_2}
\end{aligned}
\end{equation}
where $\widehat{\mathbf{N}}_1$ and $\widehat{\mathbf{N}}_2$ are the remaining unwanted part from the equations that include the \ac{AWGN}. These values are assumed to be small compared to the desired part of the equation, therefore, the rank of the channel matrix remains the same.
Since $\mathbf{U}$ is unitary, $(\mathbf{U}^{i})^H\mathbf{U}^i = \mathbf{I}$, and $\mathbf{X}$ is semi-unitary. Then, by substituting \eqref{equ:Channel-est-sing} in \eqref{equ:RX2} and \eqref{equ:RX2_2}, we get
\begin{equation}
\begin{aligned}
   \mathbf{Y}\mathbf{Y}^H &= \mathbf{u}_1^i(\lambda_1^i)^2(\mathbf{u}_1^{i})^H + \widehat{\mathbf{N}}_1, \\
     \mathbf{X} \mathbf{Y}^H\mathbf{Y}\mathbf{X}^H &= \mathbf{v}_1^{i}(\lambda_1^i)^2(\mathbf{v}_1^{i})^H + \widehat{\mathbf{N}}_2.
    \label{equ:RX3}
\end{aligned}
\end{equation}
The result emphasizes that taking the \ac{EVD} of \eqref{equ:RX3} gives directly $\mathbf{u}_1^{i}$, $\mathbf{v}_1^{i}$, and $\lambda^i_1$, and consequently $\mathbf{\Tilde{g}}^{i} = (\mathbf{v}_1^{i})^H$ and $\mathbf{\Tilde{h}}^{i}= \lambda_1\mathbf{u}_1^{i}e^{-j\theta^i}$ can be estimated, the term $e^{-j\theta^i}$ is added to cancel the phase shift induced by the RIS. 

The \ac{EVD} problem to estimate $\mathbf{\Tilde{g}}^{i}$ and $\mathbf{\Tilde{h}}^{i}$ is equivalent to the non-iterative least square algorithm in \cite{nonittLS} introduced to solve the least square problems $\underset{\mathbf{\Tilde{h}}^{i}}{\operatorname{min}}\quad \|\mathbf{Y}\mathbf{Y}^H - \mathbf{\Tilde{h}}^{i} (\mathbf{\Tilde{h}}^{i})^H \|^2_F$ and $\underset{\mathbf{\Tilde{g}}^{i}}{\operatorname{min}}\quad \|\left(\mathbf{X} \mathbf{Y}^H\mathbf{Y}\mathbf{X}^H\right)/(\lambda_1^i)^2  - (\mathbf{\Tilde{g}}^{i})^H \mathbf{\Tilde{g}}^{i} \|^2_F$ to get $\mathbf{\Tilde{h}}^{i}$, $\lambda_1^i=\|\mathbf{\Tilde{h}}^{i}\|$, and $\mathbf{\Tilde{g}}^{i}$, respectively.

\subsection{Channel Estimation for the Whole \ac{RIS}} \label{subsection:multi-element}
Following the previous subsection, the channel estimation algorithm can be generalized for $N$ \ac{RIS} elements and can be applied for all \ac{RIS}-\ac{MIMO} systems. However, this generalization is not straightforward, and there is a constraint that must be taken into consideration. From \eqref{equ:EVD} and \eqref{equ:single-element-matrix}, it is found that 
\begin{equation}
    \mathbf{H}_{T} =\sum_{m=1}^{\min(N_t,N_r)} \mathbf{u}_m\lambda_m\mathbf{v}_m^*= \sum_{i=1}^N \mathbf{h}^i\sigma_{scs}^i e^{j\theta_i}\mathbf{g}^i.
    \label{equ:constraint-nt}
\end{equation}
This indicates that the \ac{EVD} channel estimation problem holds only if $\min(N_t,N_r) \geq N$. Hence, \ac{RIS} elements are divided into subgroups $N_{sub}$ with $\min(N_t,N_r)$ elements in each subgroup. Thus, the time overhead of the total estimation procedure is $N_t \times \lceil \frac{N}{\min(N_t,N_r)} \rceil$, where $\lceil \cdot \rceil$ returns the smallest integer value that is bigger than or equal to a number.
\footnote{\textbf{Remark 2.}\textit{ Since the transmission is operating in \ac{TDD} mode, channel estimation in \ac{RIS}-assisted networks is reciprocal in both uplink and downlink, this was shown in the work of Molisch \cite{Molisch}, where it is emphasized that the double-directional channel (\ac{RIS}-channel) is reciprocal.} }
The subgrouping is recommended to include non-adjacent RIS elements so that the channels estimated are uncorrelated with higher probability.
For the $n$-th activated subgroup, the same steps in Subsection \ref{subsection:single-RIS} can be applied to get
\begin{equation}
    \mathbf{Y}^n(\mathbf{Y}^n)^H = \mathbf{U}^{n}(\Lambda^n)^2(\mathbf{U}^{n})^H + \widehat{\mathbf{N}}_1^n,
    \label{equ:RX33_1}
\end{equation}
\begin{equation}
   \mathbf{X}(\mathbf{Y}^n)^H\mathbf{Y}^n\mathbf{X}^H =\mathbf{V}^{n}(\Lambda^n)^2(\mathbf{V}^{n})^H + \widehat{\mathbf{N}}_2^n.
   \label{equ:RX33_2}
\end{equation}

% ----------------------------
% \begin{algorithm} [t!]
% \DontPrintSemicolon
%   \KwInput{Received noisy signal ${\mathbf{Y}}$ in \eqref{equ:Rx}}
%   \KwOutput{$\mathbf{\Tilde{G}}$,$\mathbf{\Tilde{H}}$}
%   \textbf{Estimate} the total channel $\mathbf{\Tilde{H}}_T$.\;
 
%   { \eIf{$N_r>N_t$} {
%   \For{n=1,2,$\dots$,$\lceil \frac{N}{N_r} \rceil$}
%       { \textbf{Obtain} the \ac{EVD} of $\mathbf{X}(\mathbf{Y}^n)^H\mathbf{Y}^n\mathbf{X}^H$. \;
%         \textbf{Compute} $\mathbf{\Tilde{G}}^{n} = \mathbf{V}^{n}$.\;
%         }
%          \textbf{Compute} $\mathbf{\Tilde{G}} = \sum _{n}\mathbf{\Tilde{G}}^n$.\;
%          \textbf{Estimate} $\mathbf{\Tilde{H}}=\mathbf{\Tilde{H}}_{T} \mathbf{\Tilde{G}}^H(\mathbf{\Tilde{G}}\mathbf{\Tilde{G}}^H)^{-1}\mathbf{\Theta}^{-1}$ .\;
%         }
%         { 
%         \For{n=1,2,$\dots$,$\lceil \frac{N}{N_t} \rceil$}
%         {\textbf{Obtain} the \ac{EVD} of $\mathbf{Y}^n(\mathbf{Y}^n)^H$. \;
%         \textbf{Compute} $\mathbf{\Tilde{H}}^{n} = \Lambda^n\mathbf{U}^{n}(\mathbf{\Theta}^n)^{-1}$.\;
%           }
%         \textbf{Compute} $\mathbf{\Tilde{H}} = \sum _{n}\mathbf{\Tilde{H}}^n$.\;
%         \textbf{Estimate} $\mathbf{\Tilde{G}}= \mathbf{\Theta}^{-1}(\mathbf{\Tilde{H}}^H\mathbf{\Tilde{H}})^{-1}\mathbf{\Tilde{H}}^H\mathbf{\Tilde{H}}_{T}$.
%           }
%           }
% \caption{The proposed channel estimation.}
% \label{algorithm:estimation}
% \end{algorithm}
% ----------------------------------
\begin{figure}[t!]
    \centering
    \includegraphics[scale=0.48]{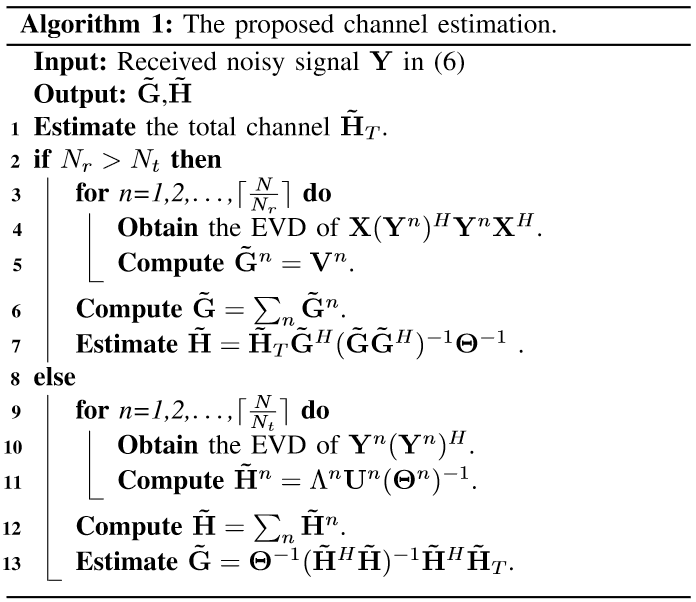}
\end{figure}

Since $\min(N_t,N_r)$ \ac{RIS} antenna elements are activated, the estimated channel for the $n$-th subgroup would be full-rank channel, and the channels can be estimated by getting the EVD of both \eqref{equ:RX33_1} and \eqref{equ:RX33_2} to get the estimated channels $\mathbf{\Tilde{H}}^{n}$ and $\mathbf{\Tilde{G}}^{n}$.
%Noting that the matrices in \eqref{equ:RX33_1} and \eqref{equ:RX33_2} are square matrices and hence the EVD becomes eigen decomposition.
Therefore, decomposing \eqref{equ:RX33_1} gives $\Lambda^n$ and $\mathbf{U}^{n}$ that are used to find $\mathbf{\Tilde{H}}^{n} = \Lambda^n\mathbf{U}^{n}(\mathbf{\Theta}^n)^{-1}$, while decomposing \eqref{equ:RX33_2} gives $\mathbf{V}^{n}$ which is equal to $\mathbf{\Tilde{G}}^{n}$, where $\mathbf{\Theta}^n$ is the diagonal matrix containing the phase shifts of the $n$-th RIS subgroup elements.

After $N_t\times \lceil \frac{N}{\min(N_t,N_r)} \rceil$ time slots, the total estimated \ac{MIMO} channel of the \ac{RIS} is given by
\begin{equation}
\mathbf{\Tilde{H}}_{T}= \sum_{n=1}^{\lceil \frac{N}{\min(N_t,N_r)} \rceil} \mathbf{\Tilde{H}}_{T}^n= \sum_{n=1}^{\lceil \frac{N}{\min(N_t,N_r)} \rceil} \mathbf{\Tilde{H}}^{n}\mathbf{\Theta}^n \mathbf{\Tilde{G}}^{n}.
 \label{equ:H-single}
\end{equation}

For further reduction in time overhead of the channel estimation procedure, we enhance the proposed algorithm to compute only one channel and find the other one using the total estimation of the effective channel (i.e., for $N_t>N_r$ or $N_t<N_r$ only $\mathbf{\Tilde{H}}$ or $\mathbf{\Tilde{G}}$ are computed, respectively). For instance, in case of $N_t>N_r$, $\mathbf{Y}\mathbf{Y}^H$ is calculated only, and hence channel $\mathbf{\Tilde{H}}^{n}$ is obtained for $n=[1,\dots,\lceil \frac{N}{N_t} \rceil]$ subgroups. Next, all \ac{RIS} are activated and the effective \ac{MIMO} channel $\mathbf{H}_T$ is estimated conventionally by considering the RIS as a random scatterer in the environment \cite{MIMOsyst}. Finally, $\mathbf{\Tilde{G}}$ is obtained as  
\begin{equation}
\mathbf{\Tilde{G}}= \mathbf{\Theta}^{-1}(\mathbf{\Tilde{H}}^H\mathbf{\Tilde{H}})^{-1}\mathbf{\Tilde{H}}^H\mathbf{\Tilde{H}}_{T} .
 \label{equ:H-single}
\end{equation}
This would result in further reducing the overhead to $N_t \times (\lceil \frac{N}{\max(N_t,N_r)} \rceil+1)$.
In case $N_r>N_t$, similar steps can be followed to estimate the cascaded channels.
The proposed scheme is summarized in Algorithm 1.

% ===========================================================
\section{Simulation Results} \label{section:Simulation-Results}

In this section, simulation results are provided to validate the proposed \ac{RIS}-\ac{MIMO} model and evaluate the proposed channel estimation framework.
The estimation performance is evaluated in terms of normalized mean-square-error (\ac{NMSE}) and the estimation performance results are obtained by averaging 10000 independent random channel realizations.

\begin{figure}[t]
    \centering
    \includegraphics[scale=0.39]{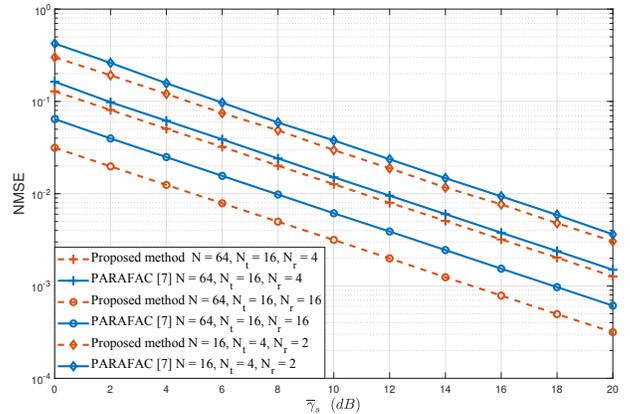}
    \caption{\ac{NMSE} performance of the effective cascaded channel by proposed channel estimation framework compared to \cite{parafac2}.}
    \label{fig:NMSE_TOT}
\end{figure}
To evaluate our proposed algorithm, we make a comparison with the first method introduced in \cite{parafac2} namely, \ac{LSKRF}. For a fair comparison, the same design requirements set in \cite{parafac2} are used to present the performance comparison between the proposed channel estimation framework and the \ac{LSKRF} method. This comparison is illustrated in Fig. \ref{fig:NMSE_TOT} under a different number of transmit/receive antennas and \ac{RIS} elements.
Fig. \ref{fig:NMSE_TOT} clearly shows that the proposed method achieves better performance compared to \ac{LSKRF} \cite{parafac2} in estimating the total effective channel $\mathbf{H}_T$. The results indicate that as the number of the receive antenna elements $N_r$ increases, the resolution of our proposed method increases, which results in enhancing the estimation performance.

\begin{figure} [h!]
    \centering
    \includegraphics[scale=0.39]{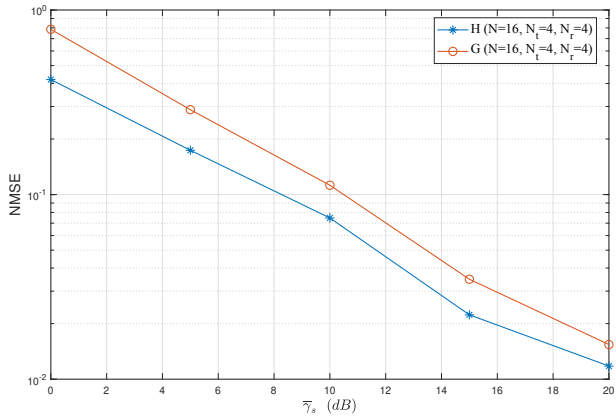}
    \caption{\ac{NMSE} performance of the cascaded channels separately when $N_t = N_r =4$ and $N=16$.}
    \label{fig:NMSE_cascaded}
\end{figure}
In order to show the main advantage of the proposed estimation algorithm, Fig. \ref{fig:NMSE_cascaded} illustrates the \ac{NMSE} performance of estimating the cascaded channel separately at $N_t=N_r=4$ and $N=16$, where the system setup is assumed to be same as in Fig. \ref{fig:NMSE_TOT}. 
The performance of each channel link is shown to be similar to conventional MIMO systems with large antenna array size \cite{MIMOsyst}, hence, controlling the channel can be feasible for each RIS-MIMO channel link separately, and the type of channel can be identified to enable more functionality such as channel tracking and precoding at the RIS level.

\begin{figure} [h] 
    \centering
    \includegraphics[scale=0.39]{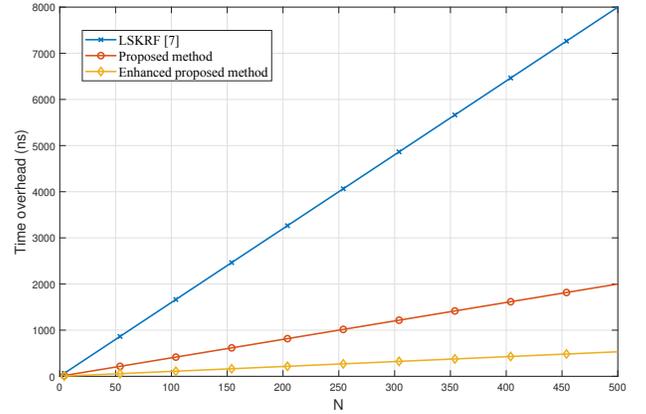}
    \caption{The time overhead of the proposed channel estimation framework compared to \cite{parafac2} for different number of \ac{RIS} elements when $N_t=16$ and $N_r=4$.}
    \label{fig:time_overhead}
\end{figure}

The time overhead of the proposed channel estimation algorithm with its enhanced version in comparison to \ac{LSKRF} method \cite{parafac2} is presented in Fig. \ref{fig:time_overhead} at different number of \ac{RIS} elements for $N_t=16$ and $N_r=4$. It is seen that \ac{LSKRF} method \cite{parafac2} requires at least $\min(N_t,N_r)$ time overhead more than our proposed algorithm for correctly estimating the total channel $\mathbf{H}_T$. For example, at $N=256$, the proposed method achieves 75\% time overhead reduction compared to \ac{LSKRF} method \cite{parafac2}, while the enhanced version of the proposed algorithm achieves more than 93\%. These reduction percentages increase as the number of the \ac{RIS} elements increases, which results in having more practical and efficient estimation method in terms of both time overhead and estimation performance. 

% ========================================================
\section{Conclusion} \label{section:Conclusion}
In this paper, the RIS-MIMO channel is modeled as a keyhole MIMO system. Based on that, we propose two novel channel estimation algorithms by unfolding the RIS-MIMO channel links and then analyzing the components of the cascaded channel applying the EVD separately. The first proposed algorithm provides $\min(N_r,N_t)$ times reduction in the system overhead compared to the conventional schemes. The second algorithm is an enhanced version of the first one to further achieve $\max(N_r,N_t)$ times reduction in the total estimation duration. As future work, the proposed method can be studied to consider different sparsity levels of the channels Tx-RIS and RIS-Rx.
% ========================================================
\section*{Acknowledgment}
This work was supported in part by the Scientific and Technological Research Council of Turkey (TUBITAK) under Grant No. 116E078.


\vfill

\end{document}